\documentclass{sig-alternate}

\usepackage{subfigure}
\usepackage{algorithm}
\usepackage{algorithmic}

\begin{document}
%
\conferenceinfo{RecSysTV}{2016 Boston, Massachusetts USA}
\CopyrightYear{2016} 
\crdata{}  

\title{A Flexible Recommendation System for Cable TV}
%
%
%
%
%

\numberofauthors{3} 
%
\author{
%
%
 \alignauthor
Diogo Gon\c{c}alves\\
       \affaddr{LaSIGE}\\
       \affaddr{Faculdade de Ci\^encias}\\
       \affaddr{Universidade de Lisboa}\\
       \affaddr{Portugal}\\
       \affaddr{dgoncalves@lasige.di.fc.ul.pt}
\alignauthor
Miguel Costa\\
       \affaddr{LaSIGE}\\
       \affaddr{Faculdade de Ci\^encias}\\
       \affaddr{Universidade de Lisboa}\\
       \affaddr{Portugal}\\
       \affaddr{migcosta@gmail.com}
\alignauthor Francisco M. Couto\\
       \affaddr{LaSIGE}\\
       \affaddr{Faculdade de Ci\^encias}\\
       \affaddr{Universidade de Lisboa}\\
       \affaddr{Portugal}\\
       \affaddr{fcouto@di.fc.ul.pt}
}
\date{30 July 1999}

\maketitle
\begin{abstract}

Recommendation systems are being explored by Cable TV operators 
to improve user satisfaction with services, such as Live TV and Video on Demand (VOD) services. 
More recently, Catch-up TV has been introduced, allowing users to watch recent broadcast content whenever they want to. These services give users a large set of options from which they can choose from, creating an information overflow problem. Thus, recommendation systems arise as essential tools to solve this problem by helping users in their selection, which increases not only user satisfaction but also user engagement and content consumption. 

In this paper we present a learning to rank approach that uses contextual information and implicit feedback to improve recommendation systems for a Cable TV operator that provides Live and Catch-up TV services. We compare our approach with existing state-of-the-art algorithms and show that
our approach is superior in accuracy, while maintaining high scores of diversity and serendipity.

\end{abstract}




\newline
\section{Introduction}

Recommendation systems are being explored by Cable TV operators to improve user satisfaction with services, such as Live TV and Video on Demand (VOD) services. 
Live TV is the most popular since it was the first offered to the public and is still the most available, where a client can watch any video content that is being broadcast live~\cite{ourpaper1}. 
VOD (video-on-demand) complements the TV offer in which a client can watch anytime any video content that was pre-recorded and made available, usually a movie or series. More recently, Catch-up TV has been introduced allowing users to watch any video content that was broadcast live up to a few days before (e.g. up to 7 days). These services give users a too large set of options from which they can choose from. 

Choosing between thousands of programs broadcast in hundreds of TV channels, plus thousands of movies and series on VOD, creates an information overflow problem. All this huge amount of possible choices turns  the search and exploration of TV guides slow and with the risk of missing the best contents for a user.
As result, the user dissatisfaction increases along with a decrease of visualization time and revenues that can even lead to churn.
If users take too much time exploring the TV guide, they might move on to another activity.  For instance, if Netflix members do not find something interesting to watch in about 60 to 90 seconds, they could lose interest and move on to something else, such as a video game, a book, or even a competing service~\cite{Gomez-Uribe2015}. Another study shows that on average users require 152 seconds, suggesting difficulties in selecting a content~\cite{Nogueira2016}.

Recommendation systems are essential tools for TV operators to help users to quickly find video contents they will like to watch, improving in this way the user satisfaction with the services provided.
Recommendation systems have proven their value in a large variety of businesses, such as video streaming (e.g. Neflix, Youtube), music streaming (e.g. Spotify, Pandora), selling products and services (e.g. Amazon, Booking), and social networks (e.g. Facebook, Linkedin).
In TV, most research on recommendation systems is focused on the VOD setting, where the Netflix prize had a great impact on improving  the state-of-the-art~\cite{amatriain2013big}. However, our previous study shows that VOD gets only 1\% of all user views~\cite{ourpaper1}. Live and Catch-up TV services are the services that users most use, but despite this fact, both are target of much less research. Live and Catch-up TV present several differences and challenges when compared to VOD:

\begin{description}
   \item dynamic program catalog: in VOD systems the catalog remains mostly static with a few programs added or removed every day at most. Collaborative-filtering algorithms tend to work well in this setting. On Live and Catch-up TV, due to the nature of broadcasting, hundreds of programs are made available and removed per hour, meaning the catalog is constantly changing. This exacerbates the new item problem significantly, since many programs do not have any user feedback associated;
   
   \item reduced user feedback: programs in VOD systems are usually available for a period of several months enabling to collect enough user feedback on them. On the other hand, Live and Catch-up TV are available during small periods. On Live TV, a program is only available during its broadcast. On Catch-up TV, a program is available for a few days after its broadcast. Thus, much less feedback is collected and each program has a small time window to be recommended.

   \item implicit user feedback: in VOD systems users give indications of their preferences by explicitly rating contents (e.g like/dislike). In Live and Catch-up TV the percentage of explicit ratings is small, which makes them unfeasible for learning user preferences. On the other hand, implicit feedback is largely available and used instead, especially, whether a user watched the content more than a percentage (e.g. watched more than 50\%).

\end{description}

In this paper, we present a recommendation system for Live and Catch-up TV based on the learning to rank framework fed by varied contextual information and implicit feedback. 
The proposed approach was evaluated with a large and real dataset extracted from a large European Cable TV operator. Results show that our approach achieves superior performance when compared to recommendation algorithms typically used in VOD.

Previous research focus mostly on accuracy (e.g. RMSE) and ranking metrics (e.g. nDCG) to evaluate the quality of recommendations~\cite{amatriain2013big}. Other metrics considering diversity and novelty are also important for users, but different users can give more value to some metrics over others. For instance, a user might prefer more diverse recommendations even if they lose accuracy. Another user might prefer the opposite. Some screens in the user interface can also be adjusted to different goals. For instance, a screen with recommendations that intend to surprise the users should show contents with higher serendipity.
Thus, we present a simple and flexible algorithm to maximize the recommendations for multiple metrics.

This paper is organized as follows: In Section 2 we describe existing recommendation approaches for each studied TV setting. Section 3 presents our learning to rank approach with contextual information. 
Section 4 details the experimental setup and Section 5 shows the results. Conclusions are presented in Section 6.

\section{Related Work}

In this section we present existing recommendation algorithms and systems applied to the TV setting.

\label{related-rec}

A common baseline algorithm is the  Popular Items~\cite{Davidson2010,yuan2015}. It returns the top-n items with the highest number of occurrences over a training period. Each user receives the same unpersonalized list of recommendations. User Popular Items is a more personalized baseline that counts the user's most watched items over the training period~\cite{yuan2015}. Each user receives the top-n items with the occurrences in their watching history.

Collaborative-filtering algorithms recommend items based on the preferences of similar users. For instance, item-based collaborative-filtering relies on user ratings to calculate the similarity between items using functions, such as the cosine similarity or pearson correlation~\cite{Sarwar2001}. Then, the more similar items not rated by the user are returned considering the previous ratings given by the same user. Matrix factorization techniques support a set of collaborative-filtering algorithms where user and items latent vectors are inferred from a rating matrix. Weighted Regularized Matrix Factorization is a common matrix factorization technique that uses alternating least squares~\cite{Hu08}.

Content-based filtering algorithms do not use user ratings. They instead use items' metadata to provide similar recommendations~\cite{Pazzani2007}. A user profile is typically created by looking at the items consumed by the user, and then, the most similar items to the profile are returned. A typical implementation is to create vector representations of items based on their metadata, such as category and author. Recommendations are generated by calculating the cosine similarity between the user and the item.

\subsection{Video On Demand (VOD)}
VOD services allow the user to watch video content when they choose to, rather than waiting for a scheduled broadcast. Access to the content is given either by paying to watch a specific item or by subscribing to a catalog of items. Each program is typically available for a large time span.

Most recommendation systems research were designed with the VOD setting in mind. Datasets, such as Movielens, have supported the evaluation of many collaborative-filtering algorithms that in general present good results~\cite{harper2015movielens}. 
The Netflix prize has also helped to enhance collaborative-filtering algorithms in this area~\cite{amatriain2013big}. 

There are many examples of VOD systems.  
Bambini et al. generated VOD recommendations using an item-based collaborative-filtering, a matrix factorization collaborative-filtering and a content-based algorithms~\cite{Cremonesi2011}. It was observed that collaborative-filtering methods performed better than the content-based one.

Yu et al.~\cite{Yu2006U} presented a study of a large deployed VOD system. They studied the user behaviour when interacting with the system and found that the recommendation system influences the popularity of the items in the system. 

Research on exploring context-aware recommendations in movies has also achieved promising results by improving the recommendations accuracy~\cite{Said2010}.

\subsection{Live TV}
In TV broadcasting, programs are scheduled to be shown at a predefined time and channel which the user has to tune to watch. The catalog of programs available to the user is highly dynamic because a program can only be watched during its scheduled broadcast. Recommendation systems for Live TV have to be able to recommend programs being broadcast at the moment or to recommend scheduled programs.
Originally, recommendation systems for TV were introduced as an extension to TV guides, where people could input their favorite programs and get recommendations for programs to watch at a later date when they were broadcast. An example of such service was PTV, a website that generated personalized TV guides~\cite{Cotter2000}. 

Since the introduction of digital Set-Top Boxes (STB), cable TV operators started to provide recommendations directly onto their users' TV sets. Zimmerman et al. introduced a recommendation system that collects implicit and explicit user feedback~\cite{zimmerman2004tv}. In 2004, TiVo allowed users to receive recommendations for TV programs along with their broadcast date~\cite{ali2004tivo}. Their STBs had the feature to auto record recommended shows so the user could watch the program at a later date.

Bambini et al. provided Live TV recommendations for new programs using a content-based algorithm that uses the items' metadata to find similarities, contributing to diminish the cold start problem~\cite{Cremonesi2011}. Turrin et al. explored the user watching habits to provide Live TV recommendations that adapt to the user interests in each timeslot~\cite{turrin2014time}. They introduced the smoothed time context algorithm that uses contextual information, such as program channel and category, to return the most appropriate recommendation for each time slot.

\subsection{Catch-up TV}
Catch-up TV is a type of service that allows users to time-shift programs previously broadcast on TV and watch them at a later date. A program is typically made available for a few days on Catch-up TV after it is broadcast on TV. An example of a Catch-up TV service is the BBC iPlayer~\cite{bbciplayer}.

Not many researchers in the recommendation systems area have focused on Catch-up TV. Some research treats Catch-up TV as a part of VOD services, partly because these services typically have similar delivery infrastructures~\cite{gallo2009multimedia,yuan2015}.
Xu et al. evaluated a Catch-up TV service for TV series~\cite{Xu2013}. They developed a recommender for previously watched series based on the user watching history. They also developed a recommender for series not watched by the user, using collaborative-filtering, content-based algorithms and matrix factorization.

\subsection{Combined Recommenders}

Some studies were conducted on recommendation systems for Live TV and VOD or Live and Catch-up TV. The IPTV recommendation system developed by Bambini et al. generates recommendations for Live TV and VOD content~\cite{Cremonesi2011}. These services are compared regarding their compatibility with existing collaborative-filtering and content-based algorithms. Due to the absence of user ratings for future Live TV content, collaborative-filtering algorithms were shown to not be effective, with content-based algorithms used as an alternative.

Yuan et al.~\cite{yuan2015} developed a context-aware recommendation system for Live and Catch-up TV. They slightly improved accuracy, diversity and novelty metrics when comparing with algorithms without context.

\section{Learning to Rank Approach with Contextual information}

Previous research showed that combinations of ranking models tend to provide better results than any single model \cite{craswell05,liu2007lbd}. 
The same is valid for recommendation systems since some algorithms work better in some settings and not so well in others \cite{Cremonesi2011}. 
For instance, we can divide the programs into 3 types:
\begin{enumerate}
\item new programs never broadcast before and, thus, there is no information on user preferences about them (cold start of items);
\item programs broadcast before, but not watched by the user (e.g. a missed episode). For these programs we only have preferences from other users;
\item programs broadcast before and watched by the user. This can be new episodes or a repeated program.
\end{enumerate}

Following this division, we expect a good performance for collaborative-filtering algorithms on points 2) and 3), because they will be able to find similarities between users according to what they watched. However, collaborative-filtering will completely fail on point 1) because these similarities will not be found due to lack of user preferences.
Content-based algorithms can give good performance for 1), since the computed similarity is based on the characteristics of programs, such as the title and category. They do not need to know whether users have watched the programs. On the other hand, they tend to recommend programs with the same characteristics of the ones previously seen, thus, affecting diversity (over-specializing).
The number of times a user watched a program is a good prediction if the user will watch it again, but can be only used for point 3).
These examples serve to show that one type of algorithm do not work well in all the cases. It is therefore necessary to combine them in a way that they can complement each other.

There are several hybrid combinations that try to get the best of each algorithm and overcome the drawbacks of algorithms individually~\cite{burke2002hybrid,burke2007hybrid,Cremonesi:2011:HAR:2039320.2039325}. Most research focus on combining content-based and collaborative-filtering approaches using strategies, such as weighting the score of each algorithm, switching between algorithms according the case or even mixing the recommendation lists produced by each algorithm. However, these strategies are difficult to tune when there are many data dimensions and possible combinations.

Learning to rank (L2R) is another way to combine these algorithms aimed to provide the best ranking list according to the preferences of the users. L2R cast the generation of recommendation lists as a supervised machine learning ranking problem \cite{liu09:_l2rank_foundations}. 
In this work, L2R algorithms receive as input a training set composed by $n$ users, where each user has associated a set of programs. Each pair <user, program> contains a feature vector and a user preference according to whether the user watched the program. 
The L2R algorithms then learn ranking models by minimizing the difference between their prediction and the user preferences. Finally, each model is evaluated with a test set similar to the training set. The predictions of the model are compared with the known user preferences (\textit{ground truth}) to measure its effectiveness.
 
L2R have been extensively employed in many systems with good results, such as web search engines, ad targeting and  recommendation systems~\cite{amatriain2013big,chapelle2011yahoo,costa2014l2r}. As far as we know, this is the first time that L2R is used to provide recommendations for users of Cable TV.
We selected LambdMART~\cite{from-ranknet-to-lambdarank-to-lambdamart-an-overview} for our experiments, but other learning to rank algorithms could also be used. 
LambdMART is a Gradient Boosted Regression Trees (a.k.a. Multiple Additive Regression Trees) algorithm which is part of the solution that won the Yahoo! Learning To Rank Challenge (Track 1)~\cite{chapelle2011yahoo}.

\subsection{Recommendation Features}

There are many techniques that exploit different data and can be used to improve the recommendation lists. For instance, demographic data can be used to tune the recommendations to users' age and gender, while context-aware data can be used to tune for a time period or location.
All these data enables more personalized recommendations. 

We computed 60 representative data features and algorithms of different recommendation approaches and included them in the dataset. For simplicity, we will refer to all as features. We give an overview of the classes of features computed, where each class exploits a different assumption:

\begin{itemize}
 \item users tend to repeat the same programs \& channels: visualization
rankings and the relative visualization time of programs and channels,
per user and averaged by all users,
show which are the programs the users see most, the time users spend
watching them and the global preferences.
It is also important to know how many episodes of a program were
watched and remain to see;
\item  users tend to repeat the same categories \& subcategories of
contents: the same type of features as above, but referring to
categories (news, movies, TV series, entertainment, sports, kids,
documentaries) and subcategories (e.g. news-science, sports-soccer, movies-action).
\item  time influences user preferences: weekend information and period of
the day of when a content was broadcast and watched helps to
differentiate user preferences over the day and week.
Since users watch more contents broadcast recently, we also included
the time passed since a content was broadcast and the time passed
since the last time an episode was watched;
\item  users tend to see programs with similar textual description of their
content: similarity functions (e.g. Jaccard index, TFxIDF) between
textual metadata (e.g. title, description, actors)
estimate how similar two contents are. Stopwords were removed (e.g. the);
\item users tend to see programs with similar characteristics: content
characteristics include features such as whether it is a one shot
program or a series with several episodes,
the number of episodes, the content age and duration;
\item  similar users watch similar programs: based on the idea of
collaborative-filtering, we used algorithms such as WRMF~\cite{Hu08} and FunkSVD~\cite{FunkSVD};
\item closer periods are more important than far ones: since the user
preferences evolve over time, the last days or weeks can provide a
better estimate of user preferences.
For instance, if a user starts watching more documentaries about war in the
last week, then he will be probably interested in similar
recommendations. Hence, we computed previous features with a week granularity.
\end{itemize}

\subsection{Multi-objective Optimization}
\label{optim}

Most recommendation systems are optimized for accuracy metrics, such as nDCG, but there are many other important metrics from a user perspective. Users want to receive a list of programs that they will like to watch, but they also want diversity between these programs. On the same way, users expect novel programs and sometimes to be surprised with something completely unexpected. Our preliminary user studies showed us that despite all these goals are important for users, the importance given to each one varies per user. There is not one objective function that fits all and could be used to optimize a general ranking model. Creating a ranking model for each user is very inefficient and does not scale well. Thus, we use a different solution instead. First, we optimize the ranking model for an accuracy metric and then re-rank the list according to an objective function. This is a simple and easy to implement solution. Moreover, it provides good results and is very flexible to adjust recommendations. This solution enables even the users to change the objective function in the user interface.

We optimize to accuracy, because there is a tradeoff between accuracy and the other metrics. For instance, by improving diversity the accuracy usually drops. Then, we apply GreedyRec (Algorithm \ref{alg1}), which is a greedy algorithm that iteratively selects the program that maximizes a given objective function up to a list of size $k$.

\begin{algorithm}\raggedright
\caption{GreedyRec$(programs,objective\_function,k)$}
\label{alg1}
\small
\begin{algorithmic}[1] 
\STATE $list \Leftarrow \{\}$
\REPEAT
\STATE $programs_i \Leftarrow arg max_{i} objective\_function(list \bigcup programs_i)$ 
\STATE $list \Leftarrow list \bigcup programs_i$
\STATE $programs \Leftarrow programs \setminus programs_i$
\UNTIL{$length(programs)>0 ~\&~ length(list)<k$}
\RETURN $list$
\end{algorithmic}
\end{algorithm}

We used the following objective function as an example for our experiment:

$objective\_function(L)=$

 $\quad 0.5\cdot nDCG(L)+0.25\cdot ILD(L)+0.25\cdot MSI(L)$

$L$ is the list of programs. This function allows to optimize for 3 metrics combined: accuracy, diversity and novelty. Other objective functions can be used to optimize for other metrics.

\section{EXPERIMENTAL SETUP}

This section presents our experimental setup.
First, we give a brief description of the L2R dataset and recommendation algorithms are compared. For last, we described the evaluation methodology and metrics.

\subsection{L2R Dataset}

The L2R dataset is composed by a set of <user, program, preference,
features> quadruples, where the preference
indicates the preference degree of the user to
the program. We consider that a program was watched if the user saw more than 50\%
of the program. In this case, we assigned a user preference of 1 to
the quadruple or 0 otherwise.
We used the percentage of
visualization as implicit feedback from the users, instead of explicit feedback such as the like/dislike
given by the users,
because there are 115 times more data.
The features represent a vector of feature
values, for the
<user, program> pair. The features are described in Section 3.1.

The dataset contains a total of 83 million quadruples extracted from 12 weeks, between October and December of 2015.
These quadruples are composed by 10 thousand users randomly chosen
that watched at least 10 programs per week, and 21 thousand programs,
which are all programs available to the users during that period.
Each user watched in average 454 programs and each program was watched
by 216 users on average.

Some programs were broadcast simultaneously in multiple channels. For example, a program might be broadcast simultaneously in Standard and High Definition. We merged the views so it only counts as one program.
A program is categorized into the following categories: News, TV Series, Entertainment, Kids, Documentaries, Sports, Movies or Adults.

\subsection{Algorithms compared}

For comparison purposes, we employed 6 algorithms to produce
recommendations for our dataset:

\begin{description}
  \item[Random:] a weak baseline that produces random recommendations. This
is the closest of not having any recommendation algorithm.
  \item[Popular:] the most watched programs among all users, assuming that on
average a user is most likely to watch what most users watched.
This is the oposite of personalized recommendations, since all users
receive the same recommendations.
  \item[UserPopular:] the programs most watched by the user (i.e. a ranking of
programs by visualization time).
  \item[WRMF:] a matrix factorization technique for collaborative-filtering where user and item latent vectors are inferred from implicit feedback~\cite{Hu08}.
  \item[Content-based:] a similarity function based on the average similarity
of a content and all the programs previously seen by the user.
The similarity is measured as the sum of the cosine similarity of
TFxIDF values for the program title, description, actors and directors.
  \item[L2R:] we used LambdaMART for this experiment. We used as input the L2R dataset extracted from the TV contents and users watching behaviour.
 
\end{description}

\subsection{Evaluation Methodology \& Metrics}
Using 10 weeks of the dataset, we conducted a 5-fold cross validation to evaluate the performance of the different algorithms. Each fold used 6 consecutive weeks, by advancing one week from the previous fold (1-6; 2-7; 3-8; 4-9 and 5-10). The first 4 weeks of each fold were used for training, the second to last for test, and the last for target. 
The final results are the averages of the five tests.

We generated top-n lists of recommendations for each user using all different algorithms. The performance of each algorithm is the average on all users. We measured lists of 5 and 10 programs, because it is the typical size of recommendation lists shown in Cable TV set-top-boxes. For Live TV, a list of recommendations is generated for programs being broadcast at the times the user accessed the system. For Catch-up TV, the recommendations aimed all programs broadcast in the last 7 days.

Each algorithm was evaluated with a set of metrics that are complementary to each other, representing different goals that a recommendation system should pursue:

\begin{description}
  \item nDCG (Normalized Cumulative Discounted Gain) is an accuracy metric that gives a higher score to programs in higher ranking positions that a user watched~\cite{Jarvelin2002}. 
  \item ILD (Intra List Diversity) is a diversity metric that calculates the average distance between all pairs of programs in the recommended list~\cite{Ziegler2005}. We defined the distance function between a pair of programs as:
$$ d(i,j) = 1-(  \frac{1}{3}cat(i,j) + \frac{1}{3} subcat(i,j) + \frac{1}{3} channel(i,j)) $$
$cat(i,j)$, $subcat(i,j)$ and $channel(i,j)$ returns 1 if the pair of
programs i and j have the same category, subcategory or channel,
respectively. They return 0 otherwise.
  \item MSI (Mean Self Information) is a novelty metric based on the number of users that did not watched the program~\cite{zhou2010}.
  \item Unexpectedness is a serendipity metric that calculates how diverse the recommendation list is from the user watching history~\cite{Murakami2008}. It uses the same distance function as in ILD.
\end{description}


\section{Results}

\begin{table*}[t]
\small
\centering
\caption{Live TV results (with Live and Catch-up TV implicit feedback)}
\label{livetvResults}
\begin{tabular}{lllllllllllll}
\hline
\textbf{} & \multicolumn{2}{c}{\textbf{Accuracy}} & \multicolumn{2}{c}{\textbf{Diversity}} & \multicolumn{2}{c}{\textbf{Novelty}} & \multicolumn{2}{c}{\textbf{Serendipity}} & \multicolumn{2}{c}{\textbf{Accuracy (new)}} & \multicolumn{2}{c}{\textbf{Global}} \\
 & \multicolumn{1}{c}{\textbf{\begin{tabular}[c]{@{}c@{}}nDCG\\ @5\end{tabular}}} & \multicolumn{1}{c}{\textbf{\begin{tabular}[c]{@{}c@{}}nDCG\\ @10\end{tabular}}} & \multicolumn{1}{c}{\textbf{\begin{tabular}[c]{@{}c@{}}ILD\\ @5\end{tabular}}} & \multicolumn{1}{c}{\textbf{\begin{tabular}[c]{@{}c@{}}ILD\\ @10\end{tabular}}} & \multicolumn{1}{c}{\textbf{\begin{tabular}[c]{@{}c@{}}MSI\\ @5\end{tabular}}} & \multicolumn{1}{c}{\textbf{\begin{tabular}[c]{@{}c@{}}MSI\\ @10\end{tabular}}} & \multicolumn{1}{c}{\textbf{\begin{tabular}[c]{@{}c@{}}Seren\\ @5\end{tabular}}} & \multicolumn{1}{c}{\textbf{\begin{tabular}[c]{@{}c@{}}Seren\\ @10\end{tabular}}} & \multicolumn{1}{c}{\textbf{\begin{tabular}[c]{@{}c@{}}nDCG\\ @5\end{tabular}}} & \multicolumn{1}{c}{\textbf{\begin{tabular}[c]{@{}c@{}}nDCG\\ @10\end{tabular}}} & \multicolumn{1}{c}{\textbf{\begin{tabular}[c]{@{}c@{}}Obj\\ @5\end{tabular}}} & \multicolumn{1}{c}{\textbf{\begin{tabular}[c]{@{}c@{}}Obj\\ @10\end{tabular}}} \\ \hline
\textbf{Random} & 0.002 & 0.002 & \textbf{0.919} & \textbf{0.920} & \textbf{0.677} & \textbf{0.679} & \textbf{0.935} & \textbf{0.935} & 0.003 & 0.003 & 0.400 & 0.401 \\
\textbf{Popular} & 0.295 & 0.252 & 0.600 & 0.550 & 0.058 & 0.081 & 0.888 & 0.887 & 0.007 & 0.006 & 0.312 & 0.283 \\
\textbf{UserPopular} & 0.699 & 0.603 & 0.708 & 0.727 & 0.186 & 0.199 & 0.872 & 0.872 & 0.000 & 0.000 & 0.573 & 0.533 \\
\textbf{WRMF} & 0.318 & 0.289 & 0.534 & 0.598 & 0.129 & 0.138 & 0.862 & 0.864 & 0.020 & 0.017 & 0.325 & 0.329 \\
\textbf{Content-based} & 0.435 & 0.396 & 0.525 & 0.550 & 0.238 & 0.242 & 0.840 & 0.836 & 0.047 & 0.038 & 0.408 & 0.396 \\
\textbf{L2R} & \textbf{0.726} & \textbf{0.631} & 0.683 & 0.692 & 0.193 & 0.206 & 0.870 & 0.869 & \textbf{0.115} & \textbf{0.081} & 0.582 & 0.540 \\
\textbf{GreedyRec} & 0.524 & 0.434 & 0.735 & 0.793 & 0.701 & 0.678 & 0.843 & 0.843 & 0.030 & 0.011 & \textbf{0.621} & \textbf{0.585} \\ \hline
\end{tabular}
\end{table*}

\begin{table*}[t]
\small
\centering
\caption{Catch-up TV results (with Live and Catch-up TV implicit feedback)}
\label{catchupResults}
\begin{tabular}{lllllllllllll}
\hline
\textbf{} & \multicolumn{2}{c}{\textbf{Accuracy}} & \multicolumn{2}{c}{\textbf{Diversity}} & \multicolumn{2}{c}{\textbf{Novelty}} & \multicolumn{2}{c}{\textbf{Serendipity}} & \multicolumn{2}{c}{\textbf{Accuracy (new)}} & \multicolumn{2}{c}{\textbf{Global}} \\
 & \multicolumn{1}{c}{\textbf{\begin{tabular}[c]{@{}c@{}}nDCG\\ @5\end{tabular}}} & \multicolumn{1}{c}{\textbf{\begin{tabular}[c]{@{}c@{}}nDCG\\ @10\end{tabular}}} & \multicolumn{1}{c}{\textbf{\begin{tabular}[c]{@{}c@{}}ILD\\ @5\end{tabular}}} & \multicolumn{1}{c}{\textbf{\begin{tabular}[c]{@{}c@{}}ILD\\ @10\end{tabular}}} & \multicolumn{1}{c}{\textbf{\begin{tabular}[c]{@{}c@{}}MSI\\ @5\end{tabular}}} & \multicolumn{1}{c}{\textbf{\begin{tabular}[c]{@{}c@{}}MSI\\ @10\end{tabular}}} & \multicolumn{1}{c}{\textbf{\begin{tabular}[c]{@{}c@{}}Seren\\ @5\end{tabular}}} & \multicolumn{1}{c}{\textbf{\begin{tabular}[c]{@{}c@{}}Seren\\ @10\end{tabular}}} & \multicolumn{1}{c}{\textbf{\begin{tabular}[c]{@{}c@{}}nDCG\\ @5\end{tabular}}} & \multicolumn{1}{c}{\textbf{\begin{tabular}[c]{@{}c@{}}nDCG\\ @10\end{tabular}}} & \multicolumn{1}{c}{\textbf{\begin{tabular}[c]{@{}c@{}}Obj\\ @5\end{tabular}}} & \multicolumn{1}{c}{\textbf{\begin{tabular}[c]{@{}c@{}}Obj\\ @10\end{tabular}}} \\ \hline
\textbf{Random} & 0.002 & 0.002 & \textbf{0.919} & \textbf{0.920} & \textbf{0.677} & \textbf{0.679} & \textbf{0.935} & \textbf{0.935} & 0.001 & 0.002 & 0.400 & 0.401 \\
\textbf{Popular} & 0.293 & 0.247 & 0.600 & 0.541 & 0.058 & 0.080 & 0.888 & 0.887 & 0.021 & 0.018 & 0.311 & 0.279 \\
\textbf{UserPopular} & 0.694 & 0.600 & 0.708 & 0.727 & 0.200 & 0.213 & 0.872 & 0.872 & 0.000 & 0.000 & 0.574 & 0.535 \\
\textbf{WRMF} & 0.306 & 0.277 & 0.523 & 0.588 & 0.153 & 0.162 & 0.856 & 0.858 & 0.034 & 0.030 & 0.322 & 0.326 \\
\textbf{Content-based} & 0.345 & 0.316 & 0.503 & 0.526 & 0.346 & 0.354 & 0.837 & 0.831 & 0.059 & 0.052 & 0.385 & 0.378 \\
\textbf{L2R} & \textbf{0.726} & \textbf{0.630} & 0.683 & 0.692 & 0.199 & 0.214 & 0.870 & 0.868 & \textbf{0.110} & \textbf{0.093} & 0.583 & 0.541 \\
\textbf{GreedyRec} & 0.556 & 0.440 & 0.753 & 0.784 & 0.606 & 0.654 & 0.850 & 0.857 & 0.027 & 0.010 & \textbf{0.618} & \textbf{0.580} \\ \hline
\end{tabular}
\end{table*}

\begin{table*}[t]
\small
\centering
\caption{Catch-up TV results (with Catch-up TV implicit feedback only)}
\label{catchupOnlyResults}
\begin{tabular}{lllllllllllll}
\hline
\textbf{} & \multicolumn{2}{c}{\textbf{Accuracy}} & \multicolumn{2}{c}{\textbf{Diversity}} & \multicolumn{2}{c}{\textbf{Novelty}} & \multicolumn{2}{c}{\textbf{Serendipity}} & \multicolumn{2}{c}{\textbf{Accuracy (new)}} & \multicolumn{2}{c}{\textbf{Global}} \\
 & \multicolumn{1}{c}{\textbf{\begin{tabular}[c]{@{}c@{}}nDCG\\ @5\end{tabular}}} & \multicolumn{1}{c}{\textbf{\begin{tabular}[c]{@{}c@{}}nDCG\\ @10\end{tabular}}} & \multicolumn{1}{c}{\textbf{\begin{tabular}[c]{@{}c@{}}ILD\\ @5\end{tabular}}} & \multicolumn{1}{c}{\textbf{\begin{tabular}[c]{@{}c@{}}ILD\\ @10\end{tabular}}} & \multicolumn{1}{c}{\textbf{\begin{tabular}[c]{@{}c@{}}MSI\\ @5\end{tabular}}} & \multicolumn{1}{c}{\textbf{\begin{tabular}[c]{@{}c@{}}MSI\\ @10\end{tabular}}} & \multicolumn{1}{c}{\textbf{\begin{tabular}[c]{@{}c@{}}Seren\\ @5\end{tabular}}} & \multicolumn{1}{c}{\textbf{\begin{tabular}[c]{@{}c@{}}Seren\\ @10\end{tabular}}} & \multicolumn{1}{c}{\textbf{\begin{tabular}[c]{@{}c@{}}nDCG\\ @5\end{tabular}}} & \multicolumn{1}{c}{\textbf{\begin{tabular}[c]{@{}c@{}}nDCG\\ @10\end{tabular}}} & \multicolumn{1}{c}{\textbf{\begin{tabular}[c]{@{}c@{}}Obj\\ @5\end{tabular}}} & \multicolumn{1}{c}{\textbf{\begin{tabular}[c]{@{}c@{}}Obj\\ @10\end{tabular}}} \\ \hline
\textbf{Random} & 0.002 & 0.003 & \textbf{0.920} & \textbf{0.921} & \textbf{0.773} & \textbf{0.774} & \textbf{0.914} & \textbf{0.914} & 0.001 & 0.001 & 0.424 & 0.425 \\
\textbf{Popular} & 0.127 & 0.134 & 0.737 & 0.798 & 0.313 & 0.344 & 0.888 & 0.892 & 0.025 & 0.026 & 0.326 & 0.352 \\
\textbf{UserPopular} & 0.432 & 0.393 & 0.733 & 0.775 & 0.474 & 0.544 & 0.761 & 0.798 & 0.000 & 0.000 & 0.518 & 0.526 \\
\textbf{WRMF} & 0.073 & 0.082 & 0.496 & 0.574 & 0.388 & 0.401 & 0.842 & 0.844 & 0.027 & 0.029 & 0.258 & 0.284 \\
\textbf{Content-based} & 0.317 & 0.330 & 0.541 & 0.564 & 0.506 & 0.530 & 0.690 & 0.692 & 0.039 & 0.040 & 0.420 & 0.438 \\
\textbf{L2R} & \textbf{0.473} & \textbf{0.459} & 0.644 & 0.671 & 0.406 & 0.424 & 0.718 & 0.726 & \textbf{0.061} & \textbf{0.061} & 0.499 & 0.503 \\
\textbf{GreedyRec} & 0.332 & 0.321 & 0.835 & 0.820 & 0.661 & 0.659 & 0.822 & 0.853 & 0.011 & 0.012 & \textbf{0.540} & \textbf{0.531} \\ \hline
\end{tabular}
\end{table*}

In this section, we present the evaluation results for each algorithm applied to Live TV and Catch-up TV. Notice that the recommendations for Live TV include just the programs that were being broadcast at the time a user was watching TV, while for Catch-up TV all programs broadcast in the last 7 days were included.
We present results for the 4 metrics, for accuracy considering only programs that the user never saw, and for the objective function described in Section~\ref{optim}.

Table~\ref{livetvResults} shows the results for Live TV using Live TV and Catch-up TV implicit feedback. Our L2R approach with LambdaMART presents the best accuracy with a NDCG@5 of 0.726 and a NDCG@10 of 0.631. It has around 4 percentage points higher  than the second algorithm, which is the UserPopular. This small difference can be explained by the fact that users tend to repeat many programs in Live TV as detailed in our Cable TV characterization~\cite{ourpaper1}.
For instance, users tend to see the same news, soccer programs and soap operas. The L2R algorithms modeled that behavior and gave higher scores to programs and channels that users usually see. We can discuss whether these recommendations are useful, since the user would probably see these programs anyway. We believe they are useful, because they serve as a shortcut to programs and increase the confidence in the system~\cite{zimmerman2004tv}.
Moreover, L2R presents a NDCG@5 of 0.115 and NDCG@10 of 0.081 when considering only the new programs that were suggested and watched, contrary to the UserPopular algorithm that only suggests repeated programs. If we relax the "preference assumption" and assume that the user likes a program if watched more than 10 minutes instead of 50\%, these metrics increase to 0.296 and 0.250. These results are much superior than the ones reported by Xu et al.~\cite{Xu2013} despite not directly comparable.

Regarding diversity, novelty and serendipity, random algorithm provides the best results. On the other hand, its accuracy is close to 0. Popular presents the second worst accuracy, showing that unpersonalized recommendations, as expected, are worse than personalized.
L2R presents good diversity (ILD@5 of 0.683 and ILD@10 of 0.692) and serendipity (Seren@5 of 0.870 and Seren@10 of 0.869), but low novelty (MSI@5 of 0.193 and MSI@10 of 0.206). 
UserPopular presents similar results. Both algorithms recommend programs that are popular for most users.

The results of the GreedyRec demonstrates well that following the objective function as intended, the diversity and novelty increased in detriment of accuracy which decreased. For the objective function the GreedyRec achieved the best score with 0.621 for the top-5 and 0.585 for the top-10 recommended programs.

Table~\ref{catchupResults} presents similar results for Catch-up TV since the algorithms used exactly the same data to learn and were evaluated in the same manner (i.e. whether the program was watched up to 7 days after the broadcast), as in Live TV. The difference is that in Live TV it was recommended only programs that were being broadcast at the time a user was watching TV. In fact, we used the same algorithms with a post-filter.
Nevertheless, results suggest that the users saw mostly programs of the same periods when they usually see TV, but this requires further study.

Table~\ref{catchupOnlyResults} shows the results for Catch-up TV, but using just the implicit feedback of Catch-up TV. This is 10\% of the volume of the data extracted from Live TV. Hence, as expected, all accuracy results decreased, showing that the data from Live TV can be used to improve the results of Catch-up TV.
The same tendencies persist, with L2R being the best algorithm for accuracy, but with a larger difference between L2R and UserPopular. It is better around 4\% in NDCG@5 and 7\% in NDCG@10. L2R has also the best accuracy for recommending new programs. GreedyRec continues to present the best score for the objective function.  

Typical approaches used in VOD, such as collaborative-filtering and
content-based filtering, present around half the accuracy of L2R and
suggest almost no new programs to the user. The bias of
collaborative-filtering methods for popular programs and the
over-specialization of content-based filtering makes them bad
solutions for Live and Catch-up TV. This is in accordance with their
low diversity and novelty.\newline\newline

\section{Conclusions}

Most research on recommendation systems for TV is designed for the VOD
setting, but VOD only gets 1\% of all views of the users of a Cable TV
operator. Moreover, our results show that typical approaches used in
VOD, such as collaborative-filtering and content-based filtering, are
not so effective for Live and Catch-up TV which receive the other
99\% of views.

We presented a learning to rank approach that uses contextual
information and implicit feedback to improve recommendation systems
for a Cable TV operator that provides Live and Catch-up TV services.
The contextual information, which we describe with insights from a
previous characterization, enables to get more personalized
recommendations. The implicit feedback augments the user preferences
learned without the users having to explicitly submit this
information. We evaluated our approach against state-of-the-art
algorithms using accuracy, diversity, novelty and serendipitious
metrics. For this evaluation we used a large and real dataset
extracted from a large European Cable TV operator. Results show that
our approach is superior in accuracy and accuracy for programs never
seen before, while maintaining high scores of diversity and serendipity. Still,
recommending programs never seen before is a much harder task that requires further investigation.

A multi-objective optimization technique based on re-ranking is
proposed, enabling to quickly adjust recommendations toward a
parameterized objective function. This allows, for instance, to
provide recommendations more accurate for some users and more diverse
for others depending on their preferences. Imagine the case where the
Cable TV operator wants to increase revenue. The objective
function may favor profit.


\section{Acknowledgments}
This work was supported by FCT through funding of LaSIGE Research Unit, ref. UID/CEC/00408/2013.\newline\newline

%
\bibliographystyle{abbrv}
\bibliography{sigproc}  

\begin{thebibliography}{10}

\bibitem{FunkSVD}
Netflix update: Try this at home.
\newblock \url{http://sifter.org/~simon/journal/20061211.html}, 2006.
\newblock Accessed 2016-04-10.

\bibitem{bbciplayer}
{BBC iPlayer}.
\newblock \url{http://www.bbc.co.uk/iplayer}, 2016.
\newblock Accessed 2016-04-10.

\bibitem{ali2004tivo}
K.~Ali and W.~Van~Stam.
\newblock {TiVo: making show recommendations using a distributed collaborative
  filtering architecture}.
\newblock In {\em Proceedings of the 10th ACM SIGKDD International Conference
  on Knowledge Discovery and Data Mining}, pages 394--401, 2004.

\bibitem{amatriain2013big}
X.~Amatriain.
\newblock {Big \& personal: data and models behind Netflix recommendations}.
\newblock In {\em Proceedings of the 2nd International Workshop on Big Data,
  Streams and Heterogeneous Source Mining: Algorithms, Systems, Programming
  Models and Applications}, pages 1--6, 2013.

\bibitem{Cremonesi2011}
R.~Bambini, P.~Cremonesi, and R.~Turrin.
\newblock {A Recommender System for an IPTV Service Provider: a Real
  Large-Scale Production Environment}.
\newblock In F.~Ricci, L.~Rokach, B.~Shapira, and P.~B. Kantor, editors, {\em
  Recommender Systems Handbook}, pages 299--331. Springer US, 2011.

\bibitem{from-ranknet-to-lambdarank-to-lambdamart-an-overview}
C.~J. Burges.
\newblock {From RankNet to LambdaRank to LambdaMART: An Overview}.
\newblock Technical report, June 2010.

\bibitem{burke2002hybrid}
R.~Burke.
\newblock Hybrid recommender systems: Survey and experiments.
\newblock {\em User modeling and user-adapted interaction}, 12(4):331--370,
  2002.

\bibitem{burke2007hybrid}
R.~Burke.
\newblock Hybrid web recommender systems.
\newblock In {\em The adaptive web}, pages 377--408. Springer, 2007.

\bibitem{chapelle2011yahoo}
O.~Chapelle and Y.~Chang.
\newblock Yahoo! learning to rank challenge overview.
\newblock {\em Journal of Machine Learning Research-Proceedings Track},
  14:1--24, 2011.

\bibitem{costa2014l2r}
M.~Costa, F.~M. Couto, and M.~J. Silva.
\newblock Learning temporal-dependent ranking models.
\newblock In {\em Proceedings of the 37th Annual ACM SIGIR Conference}, 2014.

\bibitem{Cotter2000}
P.~Cotter and B.~Smyth.
\newblock {PTV: Intelligent Personalised TV Guides}.
\newblock In {\em Proceedings of the Seventeenth National Conference on
  Artificial Intelligence and Twelfth Conference on Innovative Applications of
  Artificial Intelligence}, pages 957--964, 2000.

\bibitem{craswell05}
N.~Craswell, S.~Robertson, H.~Zaragoza, and M.~Taylor.
\newblock Relevance weighting for query independent evidence.
\newblock In {\em Proceedings of the 28th Annual International ACM SIGIR
  Conference on Research and Development in Information Retrieval}, pages
  416--423, 2005.

\bibitem{Cremonesi:2011:HAR:2039320.2039325}
P.~Cremonesi, R.~Turrin, and F.~Airoldi.
\newblock Hybrid algorithms for recommending new items.
\newblock In {\em Proceedings of the 2nd International Workshop on Information
  Heterogeneity and Fusion in Recommender Systems}, HetRec '11, pages 33--40,
  2011.

\bibitem{Davidson2010}
J.~Davidson, B.~Liebald, J.~Liu, P.~Nandy, T.~Van~Vleet, U.~Gargi, S.~Gupta,
  Y.~He, M.~Lambert, B.~Livingston, and D.~Sampath.
\newblock The youtube video recommendation system.
\newblock In {\em Proceedings of the Fourth ACM Conference on Recommender
  Systems}, RecSys '10, pages 293--296. ACM, 2010.

\bibitem{gallo2009multimedia}
D.~Gallo, C.~Miers, V.~Coroama, T.~Carvalho, V.~Souza, and P.~Karlsson.
\newblock {A multimedia delivery architecture for IPTV with P2P-based
  time-shift support}.
\newblock In {\em Consumer Communications and Networking Conference, 2009},
  pages 1--2. IEEE, 2009.

\bibitem{Gomez-Uribe2015}
C.~A. Gomez-Uribe and N.~Hunt.
\newblock {The Netflix Recommender System: Algorithms, Business Value, and
  Innovation}.
\newblock {\em ACM Trans. Manage. Inf. Syst.}, 6(4):13:1--13:19, 2015.

\bibitem{ourpaper1}
D.~Gon\c{c}alves, M.~Costa, and F.~Couto.
\newblock {A Large-Scale Characterization of User Behaviour in Cable TV}.
\newblock {\em 3rd Workshop on Recommendation Systems for Television and online
  Video (RecSysTV)}, 2016.

\bibitem{harper2015movielens}
F.~M. Harper and J.~A. Konstan.
\newblock {The MovieLens Datasets: History and Context}.
\newblock {\em ACM Trans. Interact. Intell. Syst.}, 5(4):19:1--19:19, 2015.

\bibitem{Hu08}
Y.~Hu, Y.~Koren, and C.~Volinsky.
\newblock Collaborative filtering for implicit feedback datasets.
\newblock In {\em IEEE International Conference on Data Mining (ICDM 2008)},
  pages 263--272, 2008.

\bibitem{Jarvelin2002}
K.~J\"{a}rvelin and J.~Kek\"{a}l\"{a}inen.
\newblock {Cumulated Gain-based Evaluation of IR Techniques}.
\newblock {\em ACM Trans. Inf. Syst.}, 20(4):422--446, Oct. 2002.

\bibitem{liu09:_l2rank_foundations}
T.~Liu.
\newblock {\em Learning to rank for information retrieval}, volume~3 of {\em
  Foundations and Trends in Information Retrieval}.
\newblock Now Publishers Inc., 2009.

\bibitem{liu2007lbd}
T.~Liu, J.~Xu, T.~Qin, W.~Xiong, and H.~Li.
\newblock Letor: benchmark dataset for research on learning to rank for
  information retrieval.
\newblock In {\em Proc. of SIGIR 2007 Workshop on Learning to Rank for
  Information Retrieval}, 2007.

\bibitem{Murakami2008}
T.~Murakami, K.~Mori, and R.~Orihara.
\newblock {\em New Frontiers in Artificial Intelligence: JSAI 2007 Conference
  and Workshops}, chapter Metrics for Evaluating the Serendipity of
  Recommendation Lists, pages 40--46.
\newblock Springer Berlin Heidelberg, 2008.

\bibitem{Nogueira2016}
J.~Nogueira, L.~Guardalben, B.~Cardoso, and S.~Sargento.
\newblock {Catch-up TV analytics: statistical characterization and consumption
  patterns identification on a production service}.
\newblock {\em Multimedia Systems}, pages 1--19, 2016.

\bibitem{Pazzani2007}
M.~J. Pazzani and D.~Billsus.
\newblock The adaptive web.
\newblock chapter Content-based Recommendation Systems, pages 325--341.
  Springer-Verlag, Berlin, Heidelberg, 2007.

\bibitem{Said2010}
A.~Said, S.~Berkovsky, and E.~W. De~Luca.
\newblock Putting things in context: Challenge on context-aware movie
  recommendation.
\newblock In {\em Proceedings of the Workshop on Context-Aware Movie
  Recommendation}, CAMRa '10, pages 2--6, 2010.

\bibitem{Sarwar2001}
B.~Sarwar, G.~Karypis, J.~Konstan, and J.~Riedl.
\newblock Item-based collaborative filtering recommendation algorithms.
\newblock In {\em Proceedings of the 10th International Conference on World
  Wide Web}, WWW '01, pages 285--295, 2001.

\bibitem{turrin2014time}
R.~Turrin, A.~Condorelli, P.~Cremonesi, and R.~Pagano.
\newblock {Time-based TV programs prediction}.
\newblock 2014.

\bibitem{Xu2013}
M.~Xu, S.~Berkovsky, S.~Ardon, S.~Triukose, A.~Mahanti, and I.~Koprinska.
\newblock Catch-up tv recommendations: Show old favourites and find new ones.
\newblock In {\em Proceedings of the 7th ACM Conference on Recommender
  Systems}, RecSys '13, pages 285--294, 2013.

\bibitem{Yu2006U}
H.~Yu, D.~Zheng, B.~Y. Zhao, and W.~Zheng.
\newblock Understanding user behavior in large-scale video-on-demand systems.
\newblock In {\em Proceedings of the 1st ACM SIGOPS/EuroSys European Conference
  on Computer Systems 2006}, EuroSys '06, pages 333--344, 2006.

\bibitem{yuan2015}
J.~Yuan, F.~Sivrikaya, F.~Hopfgartner, A.~Lommatzsch, and M.~Mu.
\newblock {Context-aware LDA: Balancing Relevance and Diversity in TV Content
  Recommenders}.
\newblock 2015.

\bibitem{zhou2010}
T.~Zhou, Z.~Kuscsik, J.-G. Liu, M.~Medo, J.~R. Wakeling, and Y.-C. Zhang.
\newblock Solving the apparent diversity-accuracy dilemma of recommender
  systems.
\newblock {\em Proceedings of the National Academy of Sciences},
  107(10):4511--4515, 2010.

\bibitem{Ziegler2005}
C.-N. Ziegler, S.~M. McNee, J.~A. Konstan, and G.~Lausen.
\newblock Improving recommendation lists through topic diversification.
\newblock In {\em Proceedings of the 14th International Conference on World
  Wide Web}, WWW '05, pages 22--32, 2005.

\bibitem{zimmerman2004tv}
J.~Zimmerman, K.~Kauapati, A.~L. Buczak, D.~Schaffer, S.~Gutta, and J.~Martino.
\newblock {TV personalization system}.
\newblock In {\em Personalized Digital Television}, pages 27--51. Springer,
  2004.

\end{thebibliography}
%
%

\mbox{ }

\end{document}